\newif\ifanonymous
\anonymousfalse

\documentclass[acmsmall,nonacm]{acmart}

\usepackage{xcolor}
\usepackage{accents}
\usepackage{bm}
\usepackage{csquotes}
\usepackage[binary-units=true, per-mode=symbol, per-symbol=/]{siunitx}
\usepackage{numprint}
\npdecimalsign{.}
\npthousandsep{\,}
\usepackage{booktabs}
\usepackage{multirow}
\usepackage{hhline}
\usepackage{listings}
\usepackage{todonotes}
\usepackage{csquotes}

\acmJournal{CSUR}

\begin{document}

\title{Trustworthy AI Inference Systems: An Industry Research View}



\unless\ifanonymous
    
    
    \author{Rosario Cammarota}
    \email{rosario.cammarota@intel.com}
    \author{Matthias Schunter}
    \email{mts@schunter.org}
    \author{Anand Rajan}
    \email{anand.rajan@intel.com}
    \affiliation{
    \institution{Intel Labs}
    \country{USA}
    }
    
    \author{Fabian Boemer}
    \email{fabiboemer@gmail.com}
    \affiliation{
    \institution{Intel AI}
    \country{USA}
    }
    
    \author{Amos Treiber}
    \email{treiber@encrypto.cs.tu-darmstadt.de}
    \author{Thomas Schneider}
    \email{schneider@encrypto.cs.tu-darmstadt.de}
    \author{Emmanuel Stapf}
    \email{emmanuel.stapf@trust.tu-darmstadt.de}
    \author{Ahmad-Reza Sadeghi}
    \email{ahmad.sadeghi@trust.tu-darmstadt.de}
    \affiliation{
    \institution{TU Darmstadt}
    \country{Germany}
    }
    
    \author{Daniel Demmler}
    \email{daniel.demmler@uni-hamburg.de}
    \author{Joshua Stock}
    \email{joshua.stock@uni-hamburg.de}
    \affiliation{
    \institution{University of Hamburg}
    \country{Germany}
    }
    
    \author{Christian Weinert}
    \email{christian.weinert@rhul.ac.uk}
    \affiliation{
    \institution{Royal Holloway, University of London}
    \country{UK}
    }
    
    \author{\'{A}gnes Kiss}
    \email{agnes.kiss@cispa.de}
    \affiliation{
    \institution{CISPA Helmholtz Center for Information Security}
    \country{Germany}
    }
    
    \author{Huili Chen}
    \email{huc044@ucsd.edu}
    \author{Siam Umar Hussain}
    \email{siamumar@ucsd.edu}
    \author{Sadegh Riazi}
    \email{mriazi@ucsd.edu}
    \author{Farinaz Koushanfar}
    \email{farinaz@ucsd.edu}
    \author{Saransh Gupta}
    \email{sgupta@ucsd.edu}
    \author{Tajana Rosing}
    \email{tajana@ucsd.edu}
    \author{Kamalika Chaudhuri}
    \email{kamalika@cs.ucsd.edu}
    \affiliation{
    \institution{UC San Diego}
    \country{USA}
    }
    
    \author{Hamid Nejatollahi}
    \email{hnejatol@uci.edu}
    \author{Nikil Dutt}
    \email{dutt@uci.edu}
    \author{Mohsen Imani}
    \email{m.imani@uci.edu}
    \affiliation{
    \institution{UC Irvine}
    \country{USA}
    }
    
    \author{Aydin Aysu}
    \email{aaysu@ncsu.edu}
    \author{Anuj Dubey}
    \email{adubey@ncsu.edu}
    \affiliation{
    \institution{North Carolina State University}
    \country{USA}
    }
    
    \author{Kim Laine}
    \email{kim.laine@microsoft.com}
    \affiliation{
    \institution{Microsoft Research}
    \country{USA}
    }

    \author{Fateme Sadat Hosseini}
    \author{Chengmo Yang}
    \email{chengmo@udel.edu}
    \affiliation{
    \institution{University of Delaware}
    \country{USA}
    }
    
    \author{Eric Wallace}
    \author{Pamela Norton}
    \email{pamela@borsetta.io}
    \affiliation{
    \institution{Borsetta Research}
    \country{USA}
    }

    \renewcommand{\shortauthors}{Cammarota et al.}
    
    \makeatletter
    \let\@authorsaddresses\@empty
    \makeatother
\fi

\begin{abstract}
In this work, we provide an industry research view for approaching the design, deployment, and operation of trustworthy Artificial Intelligence~(AI) inference systems.
Such systems provide customers with timely, informed, and customized inferences to aid their decision, while at the same time utilizing appropriate security protection mechanisms for AI models. Additionally, such systems should also use Privacy-Enhancing Technologies~(PETs) to protect customers' data at any time.

To approach the subject, we start by introducing current trends in AI inference systems.
We continue by elaborating on the relationship between Intellectual Property~(IP) and private data protection in such systems. Regarding the protection mechanisms, we survey the security and privacy building blocks instrumental in designing, building, deploying, and operating private AI inference systems. For example, we highlight opportunities and challenges in AI systems using trusted execution environments combined with more recent advances in cryptographic techniques to protect data in use.
Finally, we outline areas of further development that require the global collective attention of industry, academia, and government researchers to sustain the operation of trustworthy AI inference systems. 
\end{abstract}

\maketitle

\section{Introduction} \label{sec:intro}
At a macroscopic level, the players in the current AI and privacy landscape are customers (including end-users), institutions, and regulators. Customers of digital systems continue maturing awareness of the risks associated with the mishandling of and the malicious access to their private information. Such risks include loss of competitiveness, financial loss, brand erosion, identity theft, and more profoundly human rights violations.
Moreover, the expectation for AI inference systems to become integral parts of modern digital systems and the lack of understanding of how some AI models operate further exacerbate customers’ concerns.
Institutions aim to capitalize on building more accurate AI inference systems from shared data belonging to multiple stakeholders for providing better services to their customers, including end users~\cite{nextgen}. The leading technology companies improve the effectiveness of their platform recommendation systems by embedding their customers’ preference patterns in sophisticated AI models~\cite{netflix}.
Such models are part of inference systems to deploy and operate in the cloud, in the core network, on devices, or a mixture thereof. Often, the raw data processed by such systems can carry private information, such as gender, location or political views, and hence \emph{requires privacy protection at any time}~\cite{gdpr}.
Simultaneously, the AI model and its coefficients (e.g., weights and biases) often require IP protection. Institutions are also aware of the regulatory and legal risks associated with mishandling private data and data breaches~\cite{bafine,fbfine}, rendering cyber insurance policies indispensable for more and more companies~\cite{cyberins}.
Regulators create and enact regulations to protect the rights of customers and customers of customers. The enforcement of such privacy-related regulations (e.g., the~European~GDPR~\cite{gdpr} in the EU, CCPA~\cite{ccpa}, and the~HIPAA~\cite{hipaa} in the US) have already resulted in substantial fines to institutions starting in~2019\footnote{As a consequence of data breaches, British Airways PLC received a fine of~GBP~183M~\cite{bafine}, Facebook received a fine of USD 5B~\cite{fbfine}, to name a few.}. In addition, the European Commission has proposed a regulatory framework~\cite{euprop} which explicitly mentions trustworthiness as one of the criteria AI systems should fulfill.

At a microscopic level, trends in AI and the crucial dimension of privacy influence the way researchers and technologists think about privacy technologies in the context of the complex interaction between customers, institutions, and regulators. Designing technologies to protect data confidentiality at any time in the data lifecycle is not a new concept.
In the 1990s, Ann Cavoukian introduced the concept and principles of~\emph{privacy by design} that requires end-to-end data protection, which includes the protection of data in use  (including the input, the intermediate data, and the output of a computation)~\cite{privacybydesign}.
Privacy-enhancing cryptographic techniques, such as secure multi-party computation or homomorphic encryption, can protect data in use in compliance with the data privacy regulations~(e.g., GDPR). However, such methods are not yet widely adopted mainly due to their programming complexity and computational overhead compared to insecure native applications. The renewed awareness of customers’ risks with data digitization and~AI, and the expectation of~AI to play in the global economy in the foreseeable future~\cite{PredictionMachines}, have accelerated industry and academic researchers’ efforts to advance privacy-enhancing cryptographic techniques towards the realization of broader market adoption and application~\cite{inpher}.

There resides complex infrastructures in-between the micro- and macroscopic levels, comprising the cloud, the end devices, and base stations, that abstract computation, storage, and communication resources. Such infrastructures represent the contact point between customers, institutions, regulations, and AI systems. The infrastructures are also where a sustainable deployment of trustworthy AI inference systems is needed to address the mainstream concerns associated with data privacy and IP protection.

\paragraph{Deployment and Operation}
Deploying and operating trustworthy AI inference systems is challenging. In its most straightforward instance, an AI inference system captures the interactions among three players: $(i)$ the customer, who owns the input data to an AI service;~$(ii)$ the service provider, who owns, deploys, and operates the AI inference service; and~$(iii)$ the infrastructure provider~(e.g., a cloud provider or a network operator).

Such parties are mutually distrustful, can collude, and can be malicious while aiming for different goals. The AI inference service, often an IP owned (either created or acquired) by the service provider, needs to be protected from the infrastructure provider and the customer. The customer's input data may carry privacy-sensitive information (e.g., in the case of medical records), and requires protection from both the service and infrastructure provider. Furthermore, if the provider experiences security breaches, the input data, the intermediate results, and the computation's output should remain confidential. Finally, the infrastructure provider requires protection from both the service provider and the customer. These aspects all require careful consideration when deploying solutions.

\paragraph{Security Mechanisms (TEE)}
Approaching the challenge of deploying and operating trustworthy AI systems in its entirety requires thinking beyond traditional security mechanisms for confidentiality, integrity, and IP protection such as a Trusted Execution Environment~(TEE)~\cite{slalom,vg,omg}.
TEEs allow protecting the service provider's data at rest with encrypted storage, input data in transit with secure channels, and the software, the intermediate data, and outputs from the infrastructure provider. However, the input, intermediate, and output data (data in use) appear in the clear in the~TEE.
Protection of data in use should be seen at its core through the lens of the social relationship governing disclosure between and among the customer and the service provider via the platforms that collect, analyze, and manipulate information for some purpose, without necessarily requiring parties to trust each other fundamentally.
Occasionally, offering TEE protection to customers' data can be sufficient, but it is not sufficient in the general case. It is paramount to provide service providers' access to customers' data for processing purposes, while not revealing the data to the processing party.

\paragraph{Advanced Cryptographic Techniques (MPC and HE)} 
To provide service providers access to customers' data, while not revealing the data to the processing party, and to protect customers' data even in the presence of data breaches at the service or infrastructure provider, technologists can resort to advanced cryptographic techniques.
For example, secure Multi-Party Computation~(MPC) and Homomorphic Encryption~(HE) enable computation on private data without ever having plaintext access. The absence of having the corresponding decryption information enhances customers' data protection even in the case of data breaches.
Fortunately, there has been an explosion of innovation in advanced cryptographic techniques for computing on encrypted data in recent years— such innovation promise to lower technology barriers currently hindering corporations in adopting the said technologies within existing workflows.
Proofs-of-concept have appeared in the context of AI inference, specifically in Deep Learning~(DL)~\cite{TIFS,chameleon,CryptoNets,CRYPTFLOW,ngraph-he,gazelle,BCDSY20,zama_concrete}.

\paragraph{Hybrid Solutions}
Hence, while it is foreseeable for such techniques to percolate into technology, technologists must bear in mind that the development, deployment, and operation of an AI inference system must protect customers' data at any time, as well as the security of the AI service and its infrastructure.
Technology providers should foresee a future in which the technology deploys symbiotic combinations of current system security practices and advanced cryptographic techniques to build secure, private, and trustworthy AI inference systems.

\section{Tail and Headwinds}
\label{sec:trends}
In this section, we highlight catalysts and design challenges for the real-world deployment of trustworthy AI inference systems. The discussion articulates the elements of an AI inference system, roughly organized in the categories application, software, infrastructure, and hardware.

\paragraph{Application}
In the application category, vision and specifically face recognition technologies are mainstream. In the private sector, one of the most impactful events for business innovation is the approval and adoption of AI for medical and diagnostic applications (e.g., the FDA green-lighting AI-as-a-medical-device~\cite{fda_ai}). 
Additionally, early commercial applications are taking off in security, retail, and consumer electronics, in which face recognition is quickly becoming a dominant form of biometric authentication.
In the public sector, however, there is strong opposition for the deployment of face recognition technologies for surveillance-related applications, as indicated by the slowdown of the US Department of Homeland Security projects~\cite{tc}.
Even Amazon, Microsoft, and IBM recently decided to limit the commercialization of their face recognition technologies due to privacy and other societal concerns~\cite{face_rec,microsoft_facerec}.
%
%
Though it is clear that the ability to compute on encrypted data should be technically sufficient to address the privacy concerns, meaningful adoption of cryptographic privacy technologies needs other developments than just technology. \emph{Other humanity disciplines} are needed to understand the interaction with humans. In particular, both customers and institutions must have a thorough understanding of privacy technologies via educational paths, and agreements on the use of such technologies via the establishment of standards and best practices. Relevant new challenges have recently been identified~\cite{agrawal2021exploring}.

\paragraph{Software}
In the software level category, the barrier to entry for the AI space is lower than ever, thanks to open-source software. Following TensorFlow and Keras, major big-tech players have made available various front-end frameworks for developers to choose from, including Apache\textsuperscript{\textregistered} MXNet. Additionally, middle-end and back-end optimizing frameworks (e.g., Apache\textsuperscript{\textregistered} TVM, Intel\textsuperscript{\texttrademark} nGraph, OpenVino\textsuperscript{\textregistered}, and PlaidML~\cite{graphcomp}) were developed to facilitate the optimization of graph computation and their mapping on a variety of hardware targets such as CPU, GPU, and FPGA. 
The software level exhibits the path of least resistance in percolating privacy technologies in~AI systems, as illustrated by the appearance of proofs-of-concepts in the literature in the past few years~\cite{eva,ngraph-he,BCDSY20,CRYPTFLOW}.  

\paragraph{Infrastructure}
In the infrastructure category, the need for real-time decision making is pushing AI closer to the edge, from the core network to demarcation points and down to end devices. This trend gives devices the ability to process information locally and respond faster. Furthermore, the data remains in proximity or mostly close to the data owner, which inherently helps with privacy. Even so, in the presence of stolen devices and data breaches, it would be desirable to lift higher the bar of data confidentiality with privacy-enhancing cryptographic techniques.

\paragraph{Hardware}
In the hardware category, many players in the semiconductor industry and startups are focused on building chips exclusively for AI workloads. It is noteworthy that semiconductor players use specialized hardware and native data type specialization to fit in hardware constraints and reach adequate performance by trading efficiency, accuracy, and speed.
However, the sway in the trade-off is domain-specific and application-dependent. For example, in the case of face recognition to unlock devices, the computation needs to fit within tenths of a millisecond. Accuracy is essential, but a delay in response hinders usability. In other cases, such as medical diagnosis, speed is relatively not critical, but a lack of accuracy can have fatal consequences for the customers, leading to a loss of customers' trust. Similarly, both high-accuracy and timeliness are required in predicting an evacuation window given an expanding fire.

Hardware-assisted technologies such as TEE have existed for a long time to protect proprietary software IPs. Solutions exist to protect AI models from the infrastructure hosting the accelerator~\cite{sgxfpga}. Introducing hardware aids for advanced cryptographic techniques remains challenging. Such techniques often rely on data types for which mainstream devices are not designed for (e.g., high-degree polynomials with large coefficients in the case of homomorphic encryption).
There is also a lack of standardization of TEE architectures making programming difficult. The design of new standardized hardware for privacy-enhancing cryptographic techniques would be a game changer~\cite{fase,heax,cryptopim}. Making such techniques inexpensive would create opportunities inconceivable with current software implementations optimized for existing hardware targets. 

\section{Privacy Technologies Penetration}
\label{sec:privacy_inference}
Including privacy technologies into different facets of AI system design via security mechanisms and advanced cryptographic techniques has been a very active field of research in recent years.
Most research focuses on machine learning (ML) tasks, providing accurate privacy-preserving classification using artificial, specifically convolutional neural networks. In Machine Learning-as-a-Service~(MLaaS), offered by all major cloud service providers, privacy technologies can allow clients to issue classification queries without revealing the potentially sensitive information that should be classified in the clear.
In some cases, both model coefficients (weights and biases) and the functional form of the service providers' model are hidden from clients. They represent intellectual property~(IP) and might contain traces of sensitive training data. In other cases, only the query to the model is protected.

\paragraph{TEE}
Trusted Execution Environment~(TEE) architectures such as Intel\textsuperscript{\textregistered} Software Guard eXtensions~(SGX), ARM\textsuperscript{\textregistered} TrustZone\textsuperscript{\textregistered}, and Sanctuary~\cite{sanctuary}, to name a few, are hardware-assisted security architectures that create a level of security and trust that goes beyond the protection capabilities of commodity Operating Systems~(OS).

Solutions that use TEEs alone~\cite{slalom,vg,omg} can protect the IP and are often the most efficient since many TEE implementations work at the native speed of the CPU. However, the software within the TEE processes the input data (or query), intermediate results, and outputs in the clear.
There exist practical cases where protection via a TEE cannot be enforced. In such cases, other techniques such as proof of ownership~\cite{deepsecure} can be used to provide evidence that a service has been counterfeited and that illegitimate uses of the service occurred.

\paragraph{MPC}
Secure multi-party computation~(MPC) protocols~\cite{yao,gmw87} are cryptographic protocols that allow multiple parties to jointly compute a publicly known function on their private inputs while revealing no information other than the result of the computation. Therefore, MPC emulates a trusted third party's behavior within a provably secure cryptographic protocol without the need to rely on such a trusted third party.

In solutions that use MPC protocols alone~(e.g., \cite{TIFS,chameleon,CRYPTFLOW}), the participants learn the architecture of the trained~AI model.
MPC solutions protect both the query to the model and the model coefficients, but not the functional form of the model. Furthermore, when the model owner does not own the computational infrastructure (e.g., the model owner serves the model via a cloud provider), the service deployment must resort to a trusted execution environment~(TEE). The TEE protects the model coefficients and allows the model owner to outsource the model coefficients to multiple non-colluding parties that run MPC~\cite{kamara2011}.
%

Hiding even the architecture of the AI model with MPC techniques is possible by using so-called universal circuits that can be programmed by the service provider to emulate any function up to a given size~\cite{valiant1976universal,kolesnikov2008practical,ICISC,KS16,GKS17,Zhao0ZL19,agks20,Liu}. Securely evaluating such a universal circuit with an MPC protocol results in private function evaluation~(PFE), where the function itself is kept private. However, universal circuits have an inevitable logarithmic overhead~\cite{valiant1976universal} and therefore are not suitable for huge functions that occur in machine learning applications.

\paragraph{HE}
Homomorphic Encryption~(HE) is a potent tool to preserve the privacy of data and users. An encryption scheme that has homomorphic properties enables computations on encrypted data. Anyone can perform computations over a ciphertext without the need to perform decryption. In practice, only limited computation on encrypted data is feasible today, creating a challenge for finding AI model architectures that are compatible with these limitations.

HE could allow IP protection inherently when using Fully Homomorphic Encryption~(FHE) constructions~\cite{gentry2009fully,bgv,fv,fhew,tfhe}. Due to the bootstrapping procedure to refresh ciphertexts noise, the parameter selection to instantiate an FHE schema can become independent from the complexity of the function to evaluate. Hence, only the party performing the inference learns (knows) the model.
If such a party is the cloud provider (e.g., AI serviced via Google's or Microsoft's infrastructure), the model can be protected with the security schemes used to protect the cloud infrastructure and its services. If the party owning the model serves the model via a cloud infrastructure, the deployment can utilize a TEE to protect the model from the cloud infrastructure. In either case, fully homomorphic encryption methods can evaluate the model in the clear on encrypted queries while allowing computation on encrypted data owned by two (or more) distrusted parties.
Industry has recently picked up on implementations and performance evaluation in this field and commercial applications are to be expected~\cite{zama_concrete,duality,inpher_pets}.
%
%
%

\paragraph{DP}
Differential privacy~(DP)~\cite{DMNS06} is a statistical notion of privacy where the goal is to protect the privacy of individuals in the training data from an adversary who sees the trained model. Privacy is guaranteed by ensuring that the participation of a single person in the training data does not change the probability of any outcome by much. This implies that an adversary who sees the output of a differentially private algorithm cannot make any inferences about a person in the training data with high confidence that they could not make if this person had not been in the training data at all. 

In machine learning, differential privacy is mostly provided by adding noise during the training process. Noise can be added in several ways -- to the data itself, to a classifier built on the data~\cite{CMS11,rubinstein2009learning}, or to a loss function or objective~\cite{CMS11}. Currently the most common method for differentially private machine learning is to add noise at each iteration of a stochastic gradient descent process during training~\cite{SCS13,BST14,abadi2016deep}. This provides good privacy, but the added noise results in a loss of statistical efficiency -- measured by model quality per sample size. Model quality can be measured by classification accuracy (for classification models) or test log-likelihood (for generative models), and usually the loss of quality due to privacy is lower when a large amount of training data is present. While much progress has been made in the past years, ensuring high statistical efficiency remains a central research problem in differentially private machine learning.

\paragraph{Hybrid Solutions}
In addition to the protection profile, there are performance considerations that call for hybrid solutions that realize privacy-, security-, and ultimately trustworthiness-preserving building blocks. A straightforward implementation of such AI tasks with one advanced cryptographic technique only will most likely  not result in practical solutions in terms of latency, throughput, and costs for computation power or network traffic in real-world applications. For example, 
in state-of-the art instantiations of Yao's garbled circuits~MPC protocol~\cite{rosulek2021three},~197 bits of data must be sent for each \emph{binary}~AND gate which is expensive for large multiplication circuits.

Hence, alternative hybrid approaches may be required to make privacy preserving~MLaaS practically viable. Examples are the development of more efficient sub-protocols for specific tasks~(e.g., matrix multiplication), data and network pre-processing, combinations of~MPC protocols with~HE or~TEEs, optimized mixing of different~MPC protocols, and targeting more realistic and practical security models and security levels.

Modern MPC protocols are a noteworthy example of hybrid protocols. Today's most efficient~MPC protocols are hybrid protocols that combine different MPC protocols and potentially also HE for different sub-tasks~\cite{BFKL09,TIFS,tasty,ABY,chameleon,ABY3,BCDSY20}. There is also compiler support for automation~\cite{hycc,DBLP:conf/ccs/IshaqMZ19,DBLP:conf/eurosp/ChandranGRST19}. This results in highly efficient protocols for private AI (e.g., for deep neural networks)~\cite{TIFS,Secureml,oblivious,gazelle,deepsecure,chameleon,xonn,riazi2019deep,BCDSY20,Delphi}, for decision trees~\cite{BFKL09,TIFS,KNLAS19}, and for new machine learning models such as Hyperdimensional computing~\cite{imani2019framework} or sum-product networks~\cite{TMWS20}. However, most of the prior art provides solutions that are not yet easy-to-use by data scientists and have mostly been demonstrated on simple deep neural networks and small to medium datasets (e.g., MNIST~\cite{MNIST} or~CIFAR-10~\cite{CIFAR-10}).
In comparison to advanced cryptographic techniques, TEE architectures provide several orders of magnitude better performance when used for protecting AI systems~\cite{slalom,sgxfpga}. Furthermore, TEE architectures can be used to alleviate overheads introduced by advanced cryptographic techniques. For example, in~\cite{wang2019scalable}, a TEE, specifically Intel\textsuperscript{\textregistered} SGX, was used to obviate the severe overhead in the software when performing bootstrapping in FHE. The usage of a TEE architecture always requires some trust in the TEE provider. However, it is worth noting that trusting the TEE provider is a practice that is well established in the semiconductor industry to protect third party IP~\cite{Cammarota18}.

\section{Ecosystem Development} \label{sec:ecosystemAI}
Although there have been significant advances from the ongoing effort of academic, industry, and government research groups, further research and development efforts are needed to realize a meaningful commercial adoption of trustworthy~AI inference systems. For example, none of the privacy-preserving techniques (building blocks) presented above are easy to program, not to mention proving the security of finalized systems.

\paragraph{Compilers} 
For MPC and HE, recently automation tools, specific compilers, and optimizers have been designed that help developers build correct and performing solutions with MPC/HE libraries. For MPC, such tools have been available for a while~\cite{fairplay,tasty,ABY,MOTION}, including projects where tools designed for hardware development have been re-purposed to optimize according to the cost models for encrypted "gates"~\cite{tinygarble,DKSSZZ17}.
For example, CHET ~\cite{chet} is a framework optimized explicitly for neural network inference. EVA~\cite{eva} builds on CHET and adds a general-purpose compiler for a specific HE scheme.
Other projects~\cite{chet,eva,E3,ABY3} provide frameworks for targeting multiple library backends, providing a high-level unified Intermediate Representation~(IR) or Application Programming Interface~(API).
While promising, these tools are far from being mature, and mostly target advanced cryptographic techniques in the same building block family (e.g., MPC or HE).

\paragraph{Complex applications}
Mapping complex applications onto advanced cryptographic techniques is still an ad hoc process and requires almost always cryptographic expertise to be done correctly. In the case of~DL workloads, much of the literature shows that solutions with advanced cryptographic techniques can be handcrafted, but only recent work has started to show the execution of full~DL workloads. Frameworks such as~Intel\textsuperscript{\textregistered} nGraph-HE~\cite{ngraph-he,ngraph-he2,BCDSY20}, \mbox{TF-Encrypted}~\cite{tfencrypted}, and~CrypTFlow~\cite{CRYPTFLOW} allow seamless execution of pre-built neural networks on HE and MPC building blocks. CryptoSPN~\cite{TMWS20} uses sum-product networks~(SPNs) for private inference and is integrated with an~ML toolchain. 
The literature has focused on cases in which the service provider is also the owner of the infrastructure. In this case, the data owner also trains and owns the model. As new business models shift towards decentralizing the roles of the various owners in the lifecycle of AI systems, designing more complex system architectures becomes necessary, specifically combining TEEs and advanced cryptographic techniques. 
By themselves, most TEEs are also not easy to program. Fortunately, research programs such as Graphene~\cite{graphene}, a joint effort between Intel and academic partners, provides a solution to program SGX enclaves seamlessly and securely.

\section{Research and Standards} \label{sec:researchedupath}
The trend of privacy research and technology development will continue to mature current technologies ready for mass deployment. We consider hybrid solutions as viable to build trustworthy AI systems from a technology standpoint. Such solutions open opportunities to sustain the required system performance while preserving security and privacy properties. With the progress made so far, there are many areas of opportunities for further innovation developments and addressing challenges for reaching meaningful adoption.

\paragraph{Decomposition}
At a fundamental level, methods appearing in the literature are tied to specific ML workloads, typically~DL prototype networks. Thus, there is a lack of a broad understanding of how to decompose an ML workload onto a set of cryptographic mechanisms for security, data privacy, performance, and energy efficiency. Filling this knowledge gap becomes essential because new ideas and advances in ML are being proposed and adopted regularly, and the fact that privacy-preserving building blocks are aggressively introduced and refined.

\paragraph{Integration}
There is a need and opportunity to increase the integration of privacy-preserving building blocks in established ML development tools such as ML frameworks and graph compilers. For example, Intel\textsuperscript{\textregistered} nGraph-HE-ABY~\cite{BCDSY20} is an industry-class and open-source framework supporting both HE and MPC. Beyond the integration of building blocks, such a (development/testing) framework should also include automatic mechanisms to map ML workloads onto combinations of privacy-preserving building blocks for data privacy (e.g., HE versus MPC variants) and performance objectives. Additionally, it should be easy to incorporate new mechanisms as advances in privacy technologies appear in the literature to easily repair or improve existing systems.

\paragraph{Deployment}
There is also a need and opportunity to develop run-time frameworks that can reason about the available resources of computation, storage, communication, and security requirements, to automatically and expediently orchestrate the workload component for efficiency. This is particularly important due to the additional resource demand imposed with the adoption of modern privacy-preserving mechanisms and their parameter selection options.

\paragraph{Benchmarks}
A fundamental lack of benchmarks leaves another gap to be filled as many experiments restrict themselves to different tasks and assumptions. Systematic comparisons like, e.g.,~\cite{haralampieva2020systematic} for a critical mass of experiments across AI and other tasks are required at a sufficient scale to draw conclusions that can drive technology and best practices development, transfer and adoption.

\paragraph{Hardware}
New hardware is also an important topic for future research. It is paramount to design a programmable processor that provides end-to-end data security and privacy. Although new processor technology has evolved to serve complex encryption tasks more efficiently, data movement costs between the processor and memory still hinder the higher efficiency of applications relying on advanced cryptographic techniques for privacy protection~\cite{imani2019floatpim,shafiee2016isaac}. To address this issue, active research in industry and academia need to focus on designing novel architectures, including but not limited to non-von Neumann compute architectures such as in-memory or near-data computing~\cite{balasubramonian2014near,cryptopim}. 

\paragraph{Standardization}
Finally, there is a need for the development of global standards and best practices such as definitions, technical foundations and application standards to facilitate the broad deployment of advanced cryptographic mechanisms~\cite{trustworthyAI,HomomorphicEncryptionSecurityStandard}. International standards work under ISO/IEC~JTC~1 Information Security includes privacy technologies for trustworthy AI systems~\cite{iso1,iso2}, and early-stage standardization on MPC and FHE techniques. 
%

%
%
\paragraph{Provenance}
Integrity and security of AI systems are required for validating privacy requirements. Immutable ledgers for data authenticity, trusted electronic supply chains, and verified neural networks for AI inference systems provide necessary trust to enable wide distribution of the technology. There is a need for active research in integrating technologies like blockchain and distributed ledger technology into AI systems to ensure provenance and integrity throughout the manufacturing and lifecycle of the AI system components and to ensure privacy preserving requirements are met~\cite{Xu19,TAME}.  

\paragraph{Fault Resiliency}
AI models are vulnerable to various types of faults, such as permanent stuck-at defects, random bit-flip, and thermal noise~\cite{8013784,klachko2019improving}, which makes reliability critical issues in AI hardware. Both device aging and malicious activities can contribute to such reliability issues. Circuit vulnerabilities can be leveraged by attackers to intentionally cause the hardware to fail under a variety of attack vectors, including row hammer, bit-flip, gradient descent, and backdoor attacks~\cite{rakin2019bit,clements2018hardware,sreedhar2012reliability,liu2017fault,clements2018backdoor,hong2019terminal}. 
Although recent work has investigated errors in AI models~\cite{chen2017accelerator,huangfu2017computation,8119491,8060317,7167198}, they mostly focus on tolerating manufacturing defects and overlook errors caused by device aging or malicious intent. Such solutions also require non-trivial retraining and data/hardware redundancy, limiting their applicability to edge devices with tight hardware budget and limited training capability.
Hence, to monitor and maintain reliability and trustworthiness of AI accelerators, it is desirable to develop self-test and self-healing techniques that integrate test, diagnosis, and recovery loops into the system. Such techniques should target both hard defects and soft errors. They should perform a non-destructive self-test periodically to detect errors and pinpoint defective cells if any. Upon detecting any error-induced accuracy loss, such techniques should offer a retraining-free self-healing process to rescue the AI model's accuracy. The interaction of such techniques and privacy technologies is an open topic of research.

\paragraph{Side-Channel Resiliency} 
Although black-box model extraction techniques~\cite{TZJRR16,jagielski20} are ever-evolving by adopting the cryptanalytic methods~\cite{carlini2020cryptanalytic}, they are still economically infeasible, since they need millions of queries to steal a model with high fidelity.  
Side-channel analysis on inference systems~\cite{ML-SCA1,ML-SCA2,ML-SCA3,ML-SCA4,ML-SCA5,ML-SCA6,ML-SCA7,ML-SCA8,ML-phy-SCA1,ML-phys-SCA2,dubey2019maskednet,dubey2020bomanet}, by contrast, can succeed with significantly fewer tests and are harder to prevent as we learned from the research on cryptographic engineering.
Side-channel analysis can steal both the input data of the customer~\cite{ML-phy-SCA1}, as well as the model of the service provider~\cite{ML-phys-SCA2}, the general architecture of the deployed AI system~\cite{ML-SCA8}, and the detailed, bit-level values of its coefficients~\cite{dubey2019maskednet}.  
The defenses against digital side-channels, such as timing side-channels, are relatively more natural to establish for specialized accelerators (as opposed to general-purpose engines) because the design tools enable cycle-accurate control and simulation. Building defenses to prevent physical side-channel leakages such as via power consumption or electromagnetic radiation is challenging and needs tuning for the specifics of the target algorithm and its underlying implementation. Research has shown the first proofs-of-concept by extending the techniques from cryptographic engineering and empirically validating the physical side-channel security~\cite{dubey2019maskednet,dubey2020bomanet}.  
Further research is needed to achieve provably-secure, composable, and low-cost solutions that can be deployed in real-world applications of trustworthy AI systems. However, such research is unlikely to occur without significant investment, given that it is still a challenge for cryptographic systems after two decades of work. One way to accelerate the process is government involvement in standardizing the defenses~\cite{brandao2019towards}.

\section{Privacy and Explainability} \label{sec:privacyexplainability}
The performance of AI models depends on the quality of the training data. Training datasets with inherent bias lead to models replicating that bias, sometimes resulting in positive feedback loops where biased decisions are made based on decisions from biased models, creating more biased data that the models are further trained on. Many ML models, such as DNNs, are poorly explainable~(i.e., it is hard to explain post-hoc what contributed to a specific incorrect or biased prediction).
It can be hard to understand how well or how poorly the model will behave on unexpected inputs. Privacy technologies can further complicate this situation: if the model owners cannot see the query or the result, they have little hope of detecting incorrect or biased predictions, possibly thwarting good intentions to produce fair and unbiased ML applications. Novel strategies, such as Hyperdimensional computing, show promise in supporting secure and private computation while being explainable and easily updateable online \cite{imani2019framework,priveHD}.
For a broader spectrum of applications, many explainable AI (XAI) tools have been developed in recent years. They can be helpful for explaining single decisions of arbitrary black-box classifiers~\cite{lime, anchors, shap} and for visualizing the general influence of single features~\cite{pdpice}. In addition, there are many domain-specific XAI tools, e.g., saliency maps for image classification~\cite{saliency, unmasking}.

\section{Conclusion} \label{sec:conclusion}
In this work, we identify vehicles to percolate privacy technologies for AI inference, such as open-source frameworks, infrastructure, and hardware design.
We foresee architectural solutions, including hybrid methods with security mechanisms and advanced cryptographic techniques as viable innovations for designing, building, deploying and operating secure AI inference systems. 
%
%
%

The challenge of enabling seamless, private, and secure access to data without disrupting the existing AI ecosystem and lifecycle is open and offers a fertile space for privacy technologies and the development of their use cases in fields such as healthcare, finance, and retail, to name a few.
Such developments can also be applied to generic private computing applications and are not limited to AI or deep learning use cases.

We hope that the attention to the field continues fostering productive partnerships in the foreseeable future to address some of the challenges above and bring a future where the use of AI technology is compliant, safe, private, and explainable to fully realize the anticipated benefits of AI to societies and humanity worldwide.

\ifanonymous\else
\begin{acks}
We are grateful to Grace Wei, Claire Vishik of Intel, and Lian Zhu of Vox.com for their invaluable feedback. 
%
This project has received funding from the European Research Council (ERC) under the European Union's Horizon 2020 research and innovation program (grant agreement No. 850990 PSOTI). It was co-funded by the Deutsche Forschungsgemeinschaft (DFG) -- SFB 1119 CROSSING/236615297 and GRK 2050 Privacy \& Trust/\linebreak[0]{}251805230, and by the German Federal Ministry of Education and Research and the Hessen State Ministry for Higher Education, Research and the Arts within ATHENE.
The work is in part supported in part by NSF under Award \#1943245 and SRC GRC Task 2908.001.
%
\end{acks}
\fi

\ifanonymous
\clearpage
\fi

\bibliographystyle{ACM-Reference-Format}
\bibliography{references}

\end{document}